\documentclass[a4paper,12pt,titlepage]{article}

\usepackage{graphics}
\usepackage{doi}
\usepackage{setspace}
\usepackage[nooneline]{caption}
\usepackage{siunitx}
\usepackage{upgreek}

\newlength\figurewidth
\parindent=0pt{}
\parskip=\baselineskip

\begin{document}

\title{Segregation and Preferential Sputtering of Cr in WCrY Smart Alloy}

\author{
\parbox{\textwidth}{\normalsize
Hans Rudolf Koslowski$^{a,\ast}$, Janina Schmitz$^{a,b}$, Christian Linsmeier$^{a}$
\\
\\
\normalsize{}
$^{a}$ \textit{Forschungszentrum J\"ulich GmbH, Institut f\"ur Energie- und Klimaforschung -- Plasmaphysik, 52428 J\"ulich, Germany}
\\
$^{b}$ \textit{Department of Applied Physics, Ghent University, 9000 Ghent, Belgium}
\\
\\
$^{\ast}$ Corresponding author, email: h.r.koslowski@fz-juelich.de
\\
\\
ORCID IDs:\\
H. R. Koslowski https://orcid.org/0000-0002-1571-6269\\
Ch. Linsmeier https://orcid.org/0000-0003-0404-7191\\
J. Schmitz https://orcid.org/0000-0001-8495-2583\\
}}

\date{16. Dec. 2019, accepted for publication in Nucl. Mater. Energy}

\maketitle{}

\abstract{
The temperature driven segregation of Cr to the surface of the tungsten-based WCrY alloy is analysed with low energy ion scattering of He$^{+}$ ions with an energy of 1 keV in the temperature range from room temperature to \SI{1000}{K}.
Due to the high surface sensitivity, these measurements probe only the composition of the outermost monolayer.
The surface concentration of Cr increases slightly when the temperature of the sample is increased up to \SI{700}{K} and exhibits a much stronger increase when the sample temperature is further raised.
The segregation enthalpy for Cr is obtained from the Langmuir-McLean relation and amounts to \SI{0.7}{eV}.
The surface concentration of Y shows a similar behaviour to the Cr concentration.
The temperature thresholds between slow and accelerated surface density increases for Cr and Y are nearly the same.
At a temperature of \SI{1000}{K} the low energy ion scattering detects almost no W on the surface.
The modified surface composition due to the segregated species, i.e. the mixed Cr/Y layer, stays stable during cool-down of the sample.
Preferential sputtering is investigated using ion bombardment of \SI{250}{eV} D atoms, resulting in an increase of the W surface density at room temperature.
This effect is counteracted at elevated temperatures where segregation replenishes the lighter elements on the surface and prevents the formation of an all-W surface layer.
The flux of segregating Cr atoms towards the surface is evaluated from the equilibrium between sputter erosion and segregation.
}
\vfill{}

\newpage
\section{Introduction}
The first wall in a future fusion power plant needs to ensure long reliable operational periods in the presence of high particle and heat loads \cite{bol02}.
Tungsten is the candidate material for the first wall of a fusion reactor \cite{neu16}.
During long-term operation of the power plant the first wall material will become activated due to neutron irradiation.
This poses a potential danger in the unlikely case of a loss-of-coolant event.
The nuclear decay heat could raise the temperature of the first wall up to \SI{1200}{\celsius} for several weeks or months \cite{mai07}.
A simultaneous air ingress would allow tungsten to form volatile WO$_{3}$ which is released into the reactor vessel.
This problem can be alleviated by the utilisation of self-passivating ternary tungsten alloys which were developed and investigated during recent years \cite{koc11}.
These alloys show under plasma exposure preferential sputtering of the lighter constituents, leaving behind an enriched tungsten layer which has superior sputter resistance on the surface.
In case of water or air ingress a protective layer containing stable oxides of the lighter alloying elements develops on the surface which suppresses further oxidation.
Because of this situation dependent behaviour of the alloy, these materials are commonly labelled as \em smart alloys\em\ \cite{lit17}.
Such alloys with various compositions are presently under development.

WCrY alloys were recently developed and tested in laboratory experiments (oxidation resistance) and in a plasma environment \cite{lit17}.
Plasma exposure of WCrY samples in the linear plasma device PSI-2 and subsequent analysis by secondary ion mass spectrometry confirm surface enrichment of W caused by preferential sputtering of Cr \cite{sch18}.
The plasma exposure in these experiments uses a pure D plasma with sample biasing between \SIrange{120}{220}{V} and is performed at temperatures between \SI{900}{K} and \SI{1000}{K}, which is a typical temperature range estimated for the first wall during operation of a fusion power plant and temperatures of the divertor tiles are probably even higher.
However, a reverse mechanism to the Cr depletion by preferential sputtering could be the thermal segregation of Cr to the surface of the sample at elevated temperatures.
This process cannot be assessed easily in the plasma exposure experiments.

In this work, both processes, the thermally activated segregation as well as the preferential sputtering by deuterium of the lighter alloy components, are investigated in an ion beam setup.
Low energy ion scattering (LEIS) is utilised to analyse the surface composition of a WCrY sample at various temperatures.
Surface composition changes due to preferential sputtering by D ions are investigated at different temperatures in order to resolve the influence of segregation at higher temperatures.

\newpage{}
\section{Experimental}
LEIS is an extremely surface sensitive method to analyse the elemental composition of the first monolayers of a sample \cite{bro07}.
In LEIS a beam of low energy (few \si{keV}) singly charged ions (usually noble gas ions) is directed toward the surface and ions scattered under a certain angle pass an energy analyser and are detected and counted \cite{smi71}.
The collision process can be described in binary collision approximation which allows straightforward to infer the mass of the atom from which the detected projectile atom has been scattered \cite{nie93}.

The used LEIS apparatus is described in detail in \cite{kos18}.
In brief, ions are produced in a Bayard-Alpert type ion source, extracted from the source volume with a voltage of about \SI{100}{V}, and accelerated by a second electrode to their final energy of $q \times \SI{1}{kV}$, where $q$ denotes the charge of the ion.
A \ang{90} magnetic sector field is used to select the mass and charge state of the probing ion beam.
Two sets of ion optical elements, each one consisting of an einzel lens and steering plates for lateral and vertical beam deflection, are mounted before and after the magnetic sector field and guide and focus the ion beam onto the sample.
The sample is mounted on a Prevac PTS 1000 RES/C-K sample holder which can be resistively heated up to \SI{1200}{K} and allows to measure the temperature of the front plate with a K-type thermocouple.
The scattered ions pass a \ang{90} spherical energy analyser and are counted with a channeltron electron multiplier.
The sample holder and the analyser can be independently turned around the scattering centre in order to adjust incident angle and scattering angle of the ions.
The setting and scanning of the voltages at the analyser electrodes as well as the ion counting is performed with a National Instruments PXI-8106 data acquisition board which is controlled by a personal computer running LabVIEW software.
Sample cleaning and sputter erosion is done with a Perkin--Elmer 04--161 sputter ion gun.

The WCrY alloy is produced by a field-assisted sintering technology \cite{lit17}.
The composition of the produced WCrY alloy is given in table \ref{tab-wcry}.
The WCrY ingot obtained after sintering is first cut by wire erosion to get a sample which measures \SI{10}{mm} x \SI{5}{mm} x \SI{1}{mm}.
The sample is manually ground to remove residual impurities from cutting.
For grinding silicon-carbide (SiC) papers with different particle sizes are used.
Afterwards a cloth with a diamond paste of particle size \SI{1}{\mu m} is applied.
Finally, the sample is polished to a mirror finish using an Active Oxide Polishing Suspension (OPS) with \SI{0.04}{\mu m} SiO$_{2}$ particles.

Prior to the LEIS measurements the sample is in-situ cleaned by sputtering with \SI{500}{eV} Kr$^{+}$ ions under normal incidence with a total fluence in the order of \SI{1e21}{m^{-2}}.
It is almost unavoidable that the cleaning procedure by sputtering will already modify the surface because the sputtering yields for the various constituents of the WCrY sample are different.
As a result the surface concentrations of the cleaned sample deviate from the bulk concentrations.
The deviation of the surface concentrations from the respective bulk concentrations can be derived from the equilibrium state which depends the sputter yields and is for a two component alloy given by \cite{pat67}
\begin{equation}
 \frac{c_{\rm{s,1}}}{c_{\rm{s,2}}} = \frac{Y_{\rm 2}}{Y_{\rm 1}} \frac{c_{\rm{b,1}}}{c_{\rm{b,2}}},
 \label{eq-patt}
\end{equation}
where $c$ denotes the surface/bulk concentration and $Y$ is the sputter yield.
Indices s and b refer to surface and bulk, respectively, and numbers label the alloy constituents.
Kr is deliberately chosen as sputter gas because the sputter yields for W and Cr at the low ion energy of \SI{500}{eV} do not differ very much.
The elemental sputtering yields of \SI{500}{eV} Kr$^{+}$ ions for pure W and Cr are \num{0.6} and \num{1.0} \cite{yam96}, respectively.
Since the difference in sputtering yields is less than a factor of two, the resultant deviation in surface densities is rather small, too.
In any case, it is anticipated that surface changes due to thermal segregation and preferential sputtering by low energy deuterons will be much larger and can easily be distinguished from the artefacts of surface cleaning.

The mass resolution of LEIS depends on the mass ratio between the probing ion and the surface atoms \cite{tag76}.
In that respect it would be beneficial to use incident ions with a higher mass, e.g. Ne$^{+}$ or Ar$^{+}$.
However, due to the larger masses of these ions they would induce stronger sputtering of the surface which might lead to unintended changes of the surface composition.
He$^{+}$ is chosen as probing ions in order to keep those effects reasonably small.
Although the mass resolution is not optimum, the amplitude of the various peaks in the scattering spectra can be well determined by the peak fitting procedure outlined below.
For the LEIS measurements of the thermally activated Cr (and Y) segregation, the energy spectra of \SI{1}{keV} He$^{+}$ ions are measured at various sample temperatures between room temperature (RT) and \SI{1000}{K}.
The ion beam hits the sample surface under an angle of \ang{20} with respect to the surface normal and reflected ions are detected at a scattering angle of $\theta = \ang{140}$.

The investigation of the W enrichment by preferential sputtering is performed using a D$_{2}^{+}$ ion beam with an energy of \SI{500}{eV}, i.e. \SI{250}{eV} per deuteron.
The ions are produced by electron impact on D$_{2}$.
According to \cite{rap65} most ionisation events result in molecular D$_{2}^{+}$ ions\footnote{The ion source is operated at a gas pressure of $\approx \SI{1e-2}{Pa}$ which gives a Knudsen number $K_{n} >> 1$. At these large free paths for electrons and ions the number of generated D$_{3}^{+}$ ions is negligible.}, but up to \SI{7}{\percent} of the ionising collisions lead to dissociation and produce D$^{+}$ ions which hit the sample at an energy of \SI{500}{eV}.
Deuteron fluxes of the order of \SI{1e18}{m^{-2}s^{-1}} are supplied by the sputter ion gun and, after integration for few hours, applied fluences amount to several \SI{e21}{m^{-2}}.
Since the used D$_{2}$ gas has \SI{99.8}{\percent} purity and the sputter ion gun has no mass selection, the gas is fed via a LN$_{2}$ cold trap in order to freeze out any contaminants.
The elemental sputtering yield of \SI{250}{eV} deuterons for Cr is about 0.04 and the yield for W is less than \num{3e-4}, i.e. lower by at least a factor of 100 \cite{sug16}.
The difference in sputter yields for the small fraction of D$^{+}$ ions with \SI{500}{eV} in the beam is still a factor of 30 to 40 and will have only weak influence on the effect of preferential sputtering.
The LEIS measurements are taken again with He$^{+}$ ions at an impact energy of \SI{1}{keV} under a scattering angle of \ang{140}.

\begin{figure}[ht]
\resizebox{\figurewidth}{!}{\includegraphics{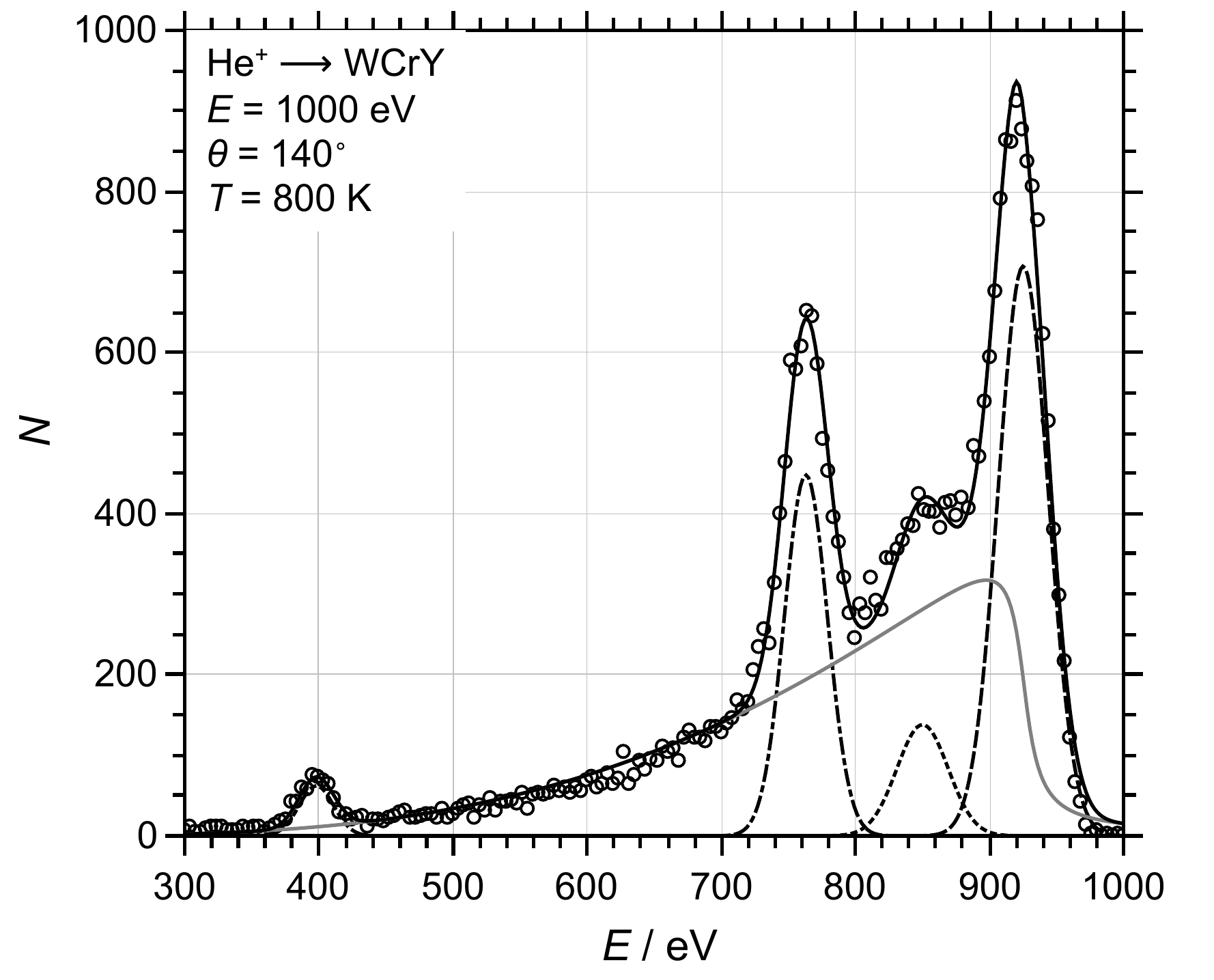}}
\caption{\label{fig-fit}\doublespacing{}
LEIS spectrum measured at a sample temperature of \SI{800}{K}.
The number of counts, $N$, is plotted versus the energy, $E$, of the respective channel.
Measured data points are drawn with open circles, the peaks in the spectrum are fitted with Gaussians: Cr (dashed-dotted line), Y(dotted line), W(dashed line).
The re-ionisation background from W is plotted with the grey line.
The small peak at \SI{400}{eV} originates from O.
The sum of all contributions is given by the full black line.}
\end{figure}

An example of a recorded LEIS spectrum is shown in figure \ref{fig-fit}.
The spectrum shows the energy distribution of the He$^{+}$ ions scattered from the WCrY sample surface while it is heated to a temperature of \SI{800}{K}.
The measuring time for this spectrum is about \SI{15}{min}.
The largest peak at the right part of the spectrum results from scattering from W atoms ($E=\SI{930}{eV}$), two smaller peaks due to scattering from Y ($E=\SI{855}{eV}$) and Cr ($E=\SI{760}{eV}$) atoms are embedded in the pronounced left wing of the W peak.
This wing results from quite strong re-ionisation of He projectiles undergoing multiple scattering events on atoms in deeper layers and therefore losing more energy.
The small peak at $E=\SI{400}{eV}$ indicates scattering from O atoms on the surface originating mainly from the yttria which is contained in the alloyed powder before sintering \cite{lit18}.
Another source could be the adsorption of O$_{2}$ from the residual gas, although the base pressure in the apparatus is below \SI{5e-8}{Pa}, limiting the oxygen exposure to less than \SI{0.3}{L h^{-1}}.
The various peaks in the spectra are fitted with Gaussians (dashed and dotted lines) and the inelastic re-ionisation background is fitted with a modified formula\footnote{The modifications are (i) the introduction of another parameter adjusting the slope of the background at the peak position and (ii) the limitation of the exponentially rising term which is kept constant for energies above the peak energy. These modifications result in considerably lower residues.} originally proposed in \cite{nel86} (grey line).
The sum of all fitted contributions (black line) shows good agreement with the measured spectrum.

The magnitude $A_{\rm i}$ of an individual peak originating from scattering at atoms of species i is given as the area under the fitted peak which is proportional to the product of peak amplitude and full width of half maximum, both obtained from the peak fitting procedure.
The uncalibrated surface concentration $c_{\rm i}$ for species i on the surface is defined as
\begin{equation}
 c_{\rm{i}} = \frac{A_{\rm{i}}}{\sum_{j} A_{\rm{j}}}
 \label{eq-surfconc}
\end{equation}
where the summation goes over all elements j on the surface.
The magnitude of an individual peak depends on the differential scattering cross section multiplied with the probability that the projectile ion is not neutralised during the collision.
This quantity might be obtained from calibration measurements using pure elemental samples of the various alloy species.
Unfortunately, such measurements are not yet available.
However, since $c_{\rm{i}}(A_{1},..,A_{n})$ defined in equation \ref{eq-surfconc} is asymptotically correct for zero ($c=0$) and full ($c=1$) coverage, and strictly monotonous in between, it fulfils the requirements to characterise the surface coverage.

\newpage{}
\section{Results and Discussion}

\begin{figure}[ht]
\resizebox{\figurewidth}{!}{\includegraphics{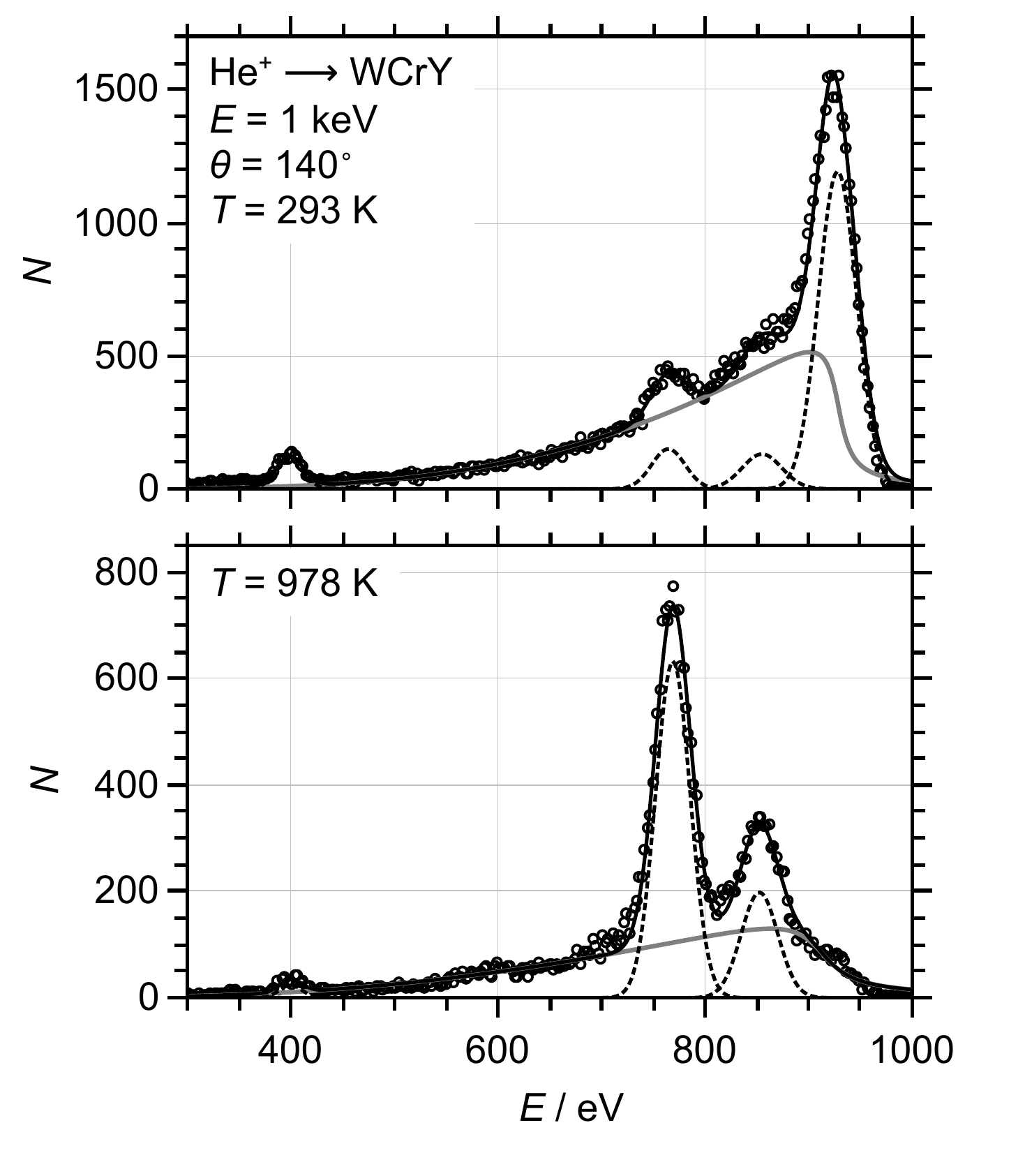}}
\caption{\label{fig-rt-1000k}\doublespacing{}
Two LEIS spectra measured at RT (top) and at $T=\SI{978}{K}$ (bottom).
In both spectra the Gaussian peak fits for W, Y, Cr, and O are shown with dotted black lines.
The re-ionisation underground is given by the full grey line and the sum of all contributions is plotted with the full black line.}
\end{figure}

Two representative LEIS spectra for the investigation of the thermally activated segregation of Cr and Y are shown in figure \ref{fig-rt-1000k}.
The spectrum at the top is measured at RT after cleaning the sample with \SI{500}{eV} Kr$^{+}$ sputtering.
The peak fitting procedure yields four distinct peaks originating from O, Cr, Y, and W (from left to right).
The spectrum shows a pronounced re-ionisation background.
The sum of the four Gaussians and the background show a good agreement with the measured data.

The LEIS spectrum at the bottom is measured at a sample temperature of \SI{978}{K}.
The Cr ($E=\SI{760}{eV}$) and Y ($E=\SI{855}{eV}$) peaks are much larger compared to the measurement at RT, indicating that segregation to the surface of both constituents has occurred.
Again, the fitting procedure yields reasonably good agreement with the measurement.
However, the W peak at $E=\SI{930}{eV}$ is suppressed, suggesting that there are almost no W atoms remaining in the top surface layer.
There is nevertheless an inelastic background contribution, extending to the original W peak position.
This shows that W atoms are still present in deeper layers and He atoms (multiply) scattered at deeper layers are re-ionised when leaving the sample.

\begin{figure}[ht]
\resizebox{\figurewidth}{!}{\includegraphics{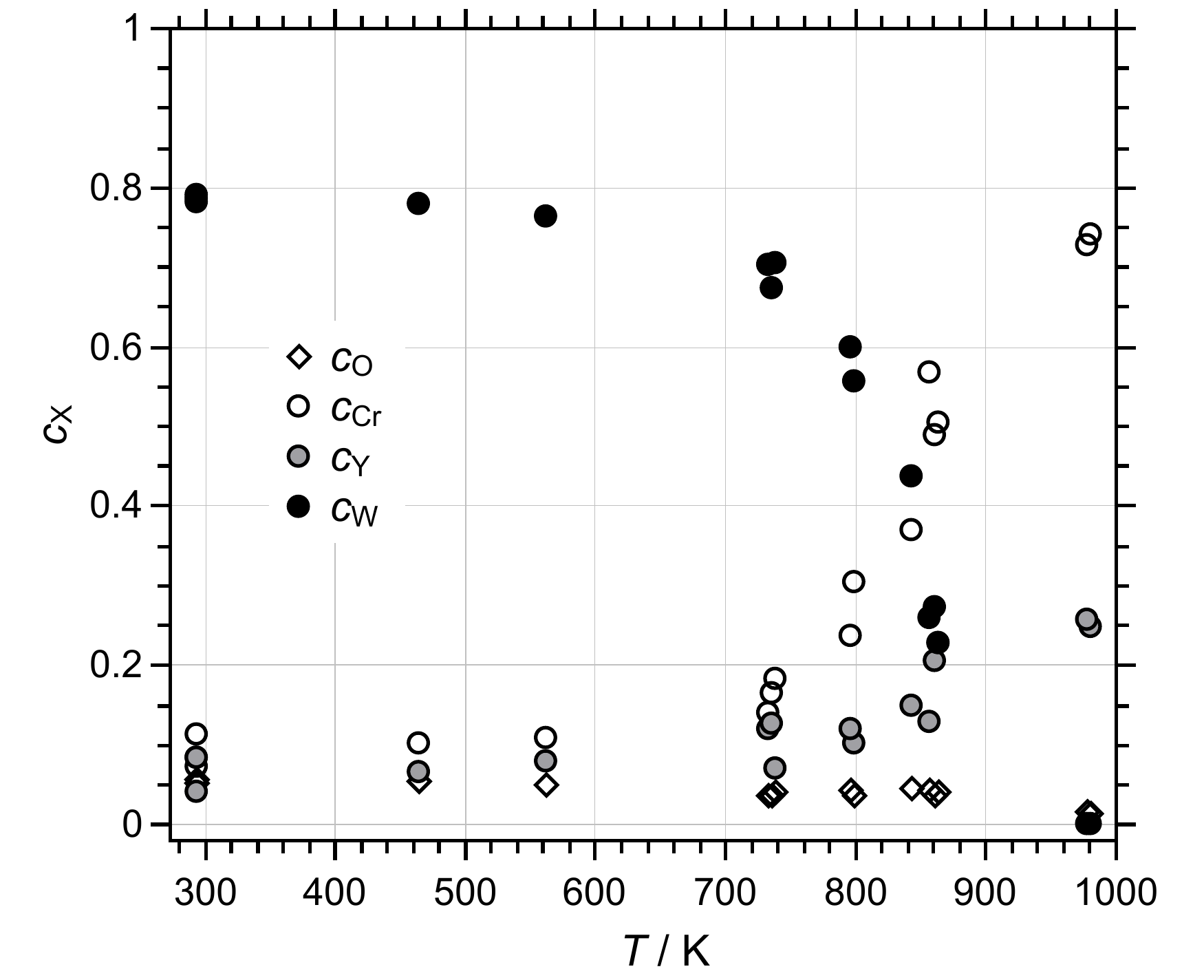}}
\caption{\label{fig-surfconc}\doublespacing{}
The fractional surface concentrations, $c_{X}$, determined from the fitted peaks in the LEIS spectra according to equation \ref{eq-surfconc} for O, Cr, Y, and W are plotted versus the sample temperature, $T$.}
\end{figure}

The measurements of the surface concentrations are performed several times.
Each run is started with sufficient sputter cleaning of the sample, thus removing the top surface layers until the LEIS spectra look similar to the spectrum shown in figure \ref{fig-rt-1000k} at the top.
Following the cleaning step the sample is heated in increasing steps up to a maximum of \SI{1000}{K}.
LEIS spectra are recorded for each temperature step and the surface concentrations of O, Cr, Y, and W are determined.
Figure \ref{fig-surfconc} depicts a compilation of all available data.
The surface concentrations change little for temperatures up to about \SI{600}{K}.
Increasing the temperature further leads to strong increases of the Cr and Y concentrations and a corresponding decrease of the W surface density.
The strong segregation occurs in the temperature range \SIrange{720}{860}{K}.
As already seen at the bottom of figure \ref{fig-rt-1000k}, no W atoms in the top surface layer are detected by LEIS at a temperature of \SI{1000}{K}.

\begin{figure}[ht]
\resizebox{\figurewidth}{!}{\includegraphics{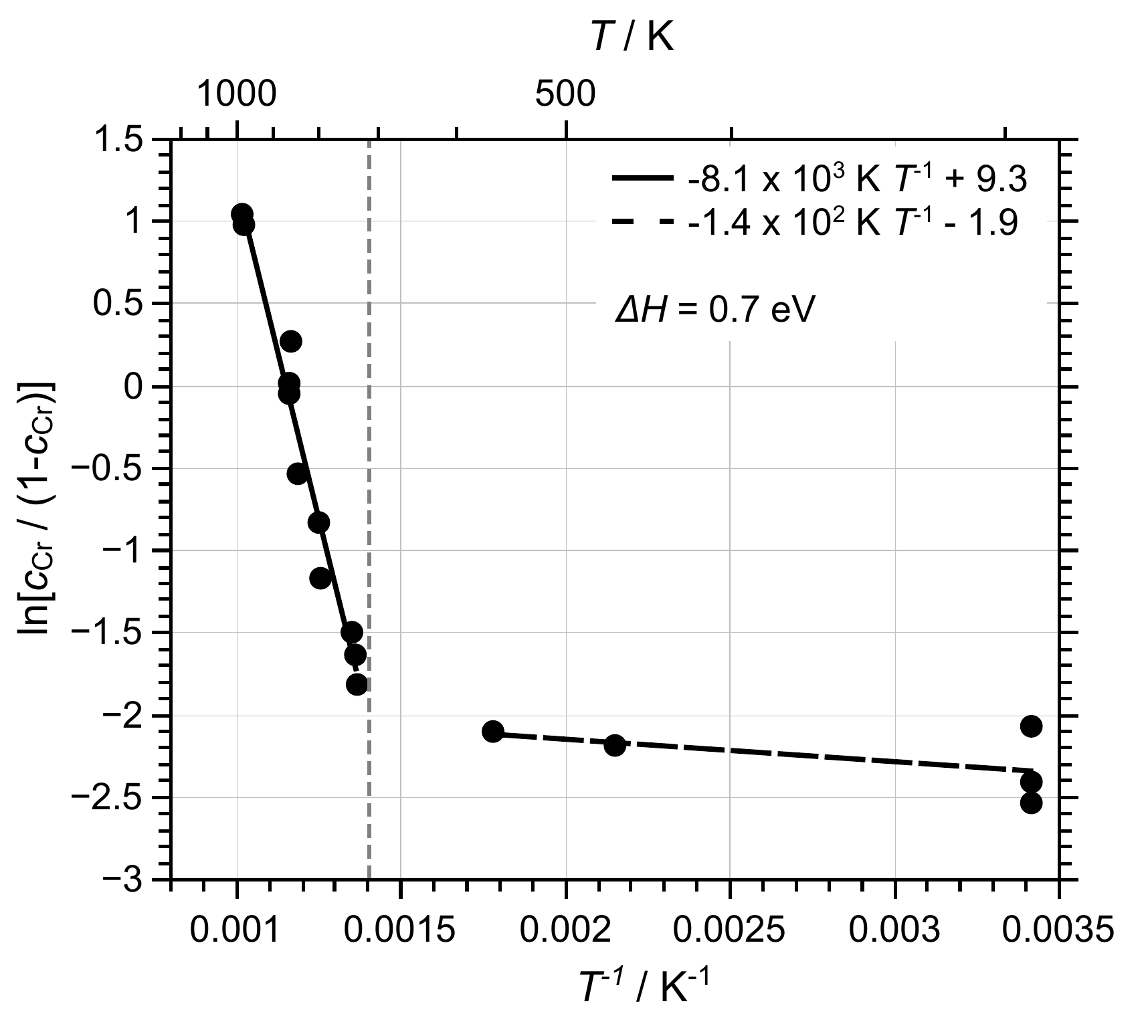}}
\caption{\label{fig-arrhenius-cr}\doublespacing{}
Arrhenius plot of the Langmuir-McLean relation (equation \ref{eq-lang-mclean}) for Cr with straight line fits for the high temperature (above \SI{700}{K}) and low temperature regions.
The logarithm of the surface coverage, $ln(\frac{c_{\rm{Cr}}^{\rm s}}{1-c_{\rm{Cr}}^{\rm s}})$, is plotted versus the reciprocal temperature, $T^{-1}$.
A temperature scale is drawn on top of the figure.
}
\end{figure}

The segregation enthalpy, $\Delta H$, can be determined from the data shown in figure \ref{fig-surfconc}.
For the analysis of the Cr segregation we follow the procedure described in \cite{bei15}.
The surface concentration of the segregating species, here Cr, is described by a Langmuir-McLean relation \cite{sea80}
\begin{equation}
 \frac{c_{\rm{Cr}}^{\rm s}}{1 - c_{\rm{Cr}}^{\rm s}} = \frac{c_{\rm{Cr}}^{\rm b}}{1 - c_{\rm{Cr}}^{\rm b}} \exp{(\frac{-\Delta G}{kT})},
\label{eq-lang-mclean}
\end{equation}
where $c_{\rm{Cr}}^{\rm s}$ denotes the fractional surface coverage of Cr, $c_{\rm{Cr}}^{\rm b}$ the fractional Cr bulk concentration, and $\Delta G$ is the Gibbs free energy of segregation.
Using $\Delta G = \Delta H - T \Delta S$, where $S$ is the entropy, and taking the logarithm of equation \ref{eq-lang-mclean} gives the following equation \cite{tau78}
\begin{equation}
 ln(\frac{c_{\rm{Cr}}^{\rm s}}{1 - c_{\rm{Cr}}^{\rm s}}) = \frac{-\Delta H}{k} \times \frac{1}{T} + const.
\label{eq-arrhenius}
\end{equation}
Figure \ref{fig-arrhenius-cr} shows the corresponding Arrhenius plot of the Cr surface concentration data.
The data points are clearly separated in two distinct regions.
At temperatures below \SI{700}{K} (right part of the figure) the Cr surface concentration changes little with increasing temperature, whereas at higher temperatures a strong increase of Cr surface coverage with increasing temperature is visible.
The lines in the figure are linear regression curves according to equation \ref{eq-arrhenius} where the data above and below \SI{700}{K} are fitted separately.
The segregation enthalpy in the high temperature region determined from the slope of the linear regression line is $\Delta H = \SI{0.7}{eV}$.
Both lines cross at a critical temperature of $T_{\rm{c,Cr}} = \SI{712}{K}$.

For the low temperature region we refrain from evaluation of a segregation enthalpy because it is not clear that the slight increase of Cr surface concentrations with rising temperature is indeed a thermal segregation as in the high temperature region.
As mentioned in the experimental section, the sputter cleaning with Kr$^{+}$ causes some preferential erosion which leads according to equation \ref{eq-patt} to a surface depletion of the species with the larger sputter yields, which is Cr.
The small increase when raising the sample temperature to \SIrange{600}{700}{K}, which is less than a factor of \num{2}, is likely due to recovery of the unperturbed surface densities due to Cr diffusion from the bulk.

\begin{figure}[ht]
\resizebox{\figurewidth}{!}{\includegraphics{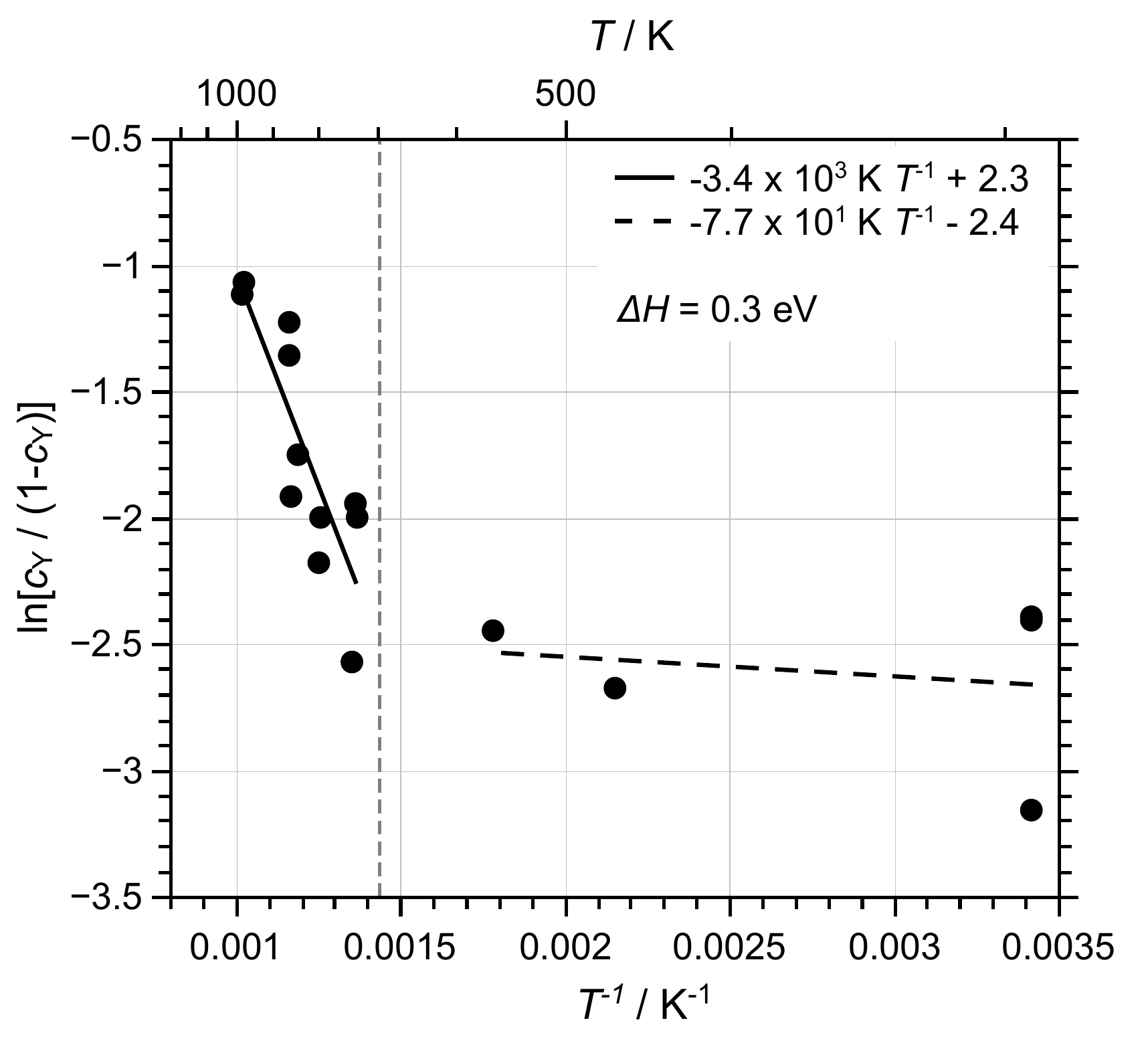}}
\caption{\label{fig-arrhenius-y}\doublespacing{}
Arrhenius plot of the Langmuir-McLean relation (equation \ref{eq-lang-mclean}) for Y with straight line fits for the high temperature (above \SI{700}{K}) and low temperature regions.
The logarithm of the surface coverage, $ln(\frac{c_{\rm{Y}}^{\rm s}}{1-c_{\rm{Y}}^{\rm s}})$, is plotted versus the reciprocal temperature, $T^{-1}$.
}
\end{figure}

The same analysis for the Y surface concentration is shown in figure \ref{fig-arrhenius-y}.
Again, the data is separated in two regions with slow and fast segregation, respectively.
The segregation enthalpy for Y is $\Delta H = \SI{0.3}{eV}$.
The critical temperature where both straight lines cross is $T_{\rm{c,Y}} = \SI{696}{K}$.
This temperature is not significantly different from the critical temperature obtained for Cr segregation.

It is obvious that for both lighter alloy constituents, Cr and Y, the segregation behaviour looks very similar.
In both cases segregation proceeds rather slowly below a critical temperature and accelerates as soon as this temperature is exceeded.
However, it is not clear what happens at the transition temperature.
Unfortunately, a ternary phase diagram of WCrY is not available in the literature.
Looking at the binary phase diagram of W-Cr, such as e.g. published in \cite{nai84}, the point of interest, $T=\SI{700}{K}$ at a Cr content of \SI{31}{\percent}, is located in the miscibility gap far from any phase boundary.
Here, the W-Cr alloy consists of a mixture of two phases called $\alpha_{1}$ and $\alpha_{2}$ which are Cr-rich and W-rich solute solutions.
The observed Cr surface enrichment might be attributed to grain boundary segregation of these phases.

\begin{figure}[ht]
\resizebox{\figurewidth}{!}{\includegraphics{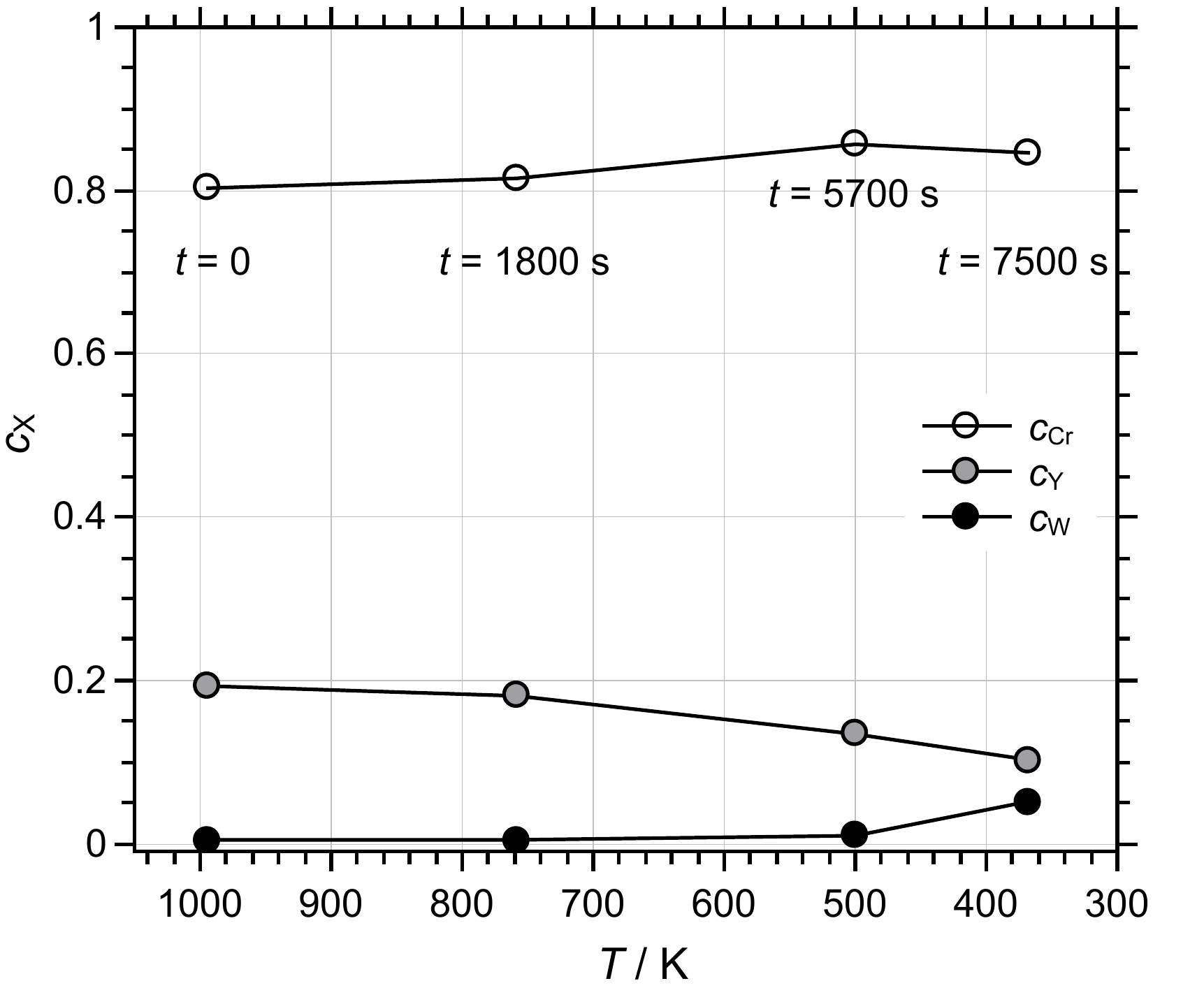}}
\caption{\label{fig-cooling}\doublespacing{}
Fractional surface concentrations, $c_{X}$, measured for various temperatures during cooling of the sample.
The measurements proceed from left to right. Note that the temperature scale, $T$, is inverted.
The times when each spectrum has been measured are indicated in the figure.}
\end{figure}

After having quantified the segregation behaviour, an interesting questions arises: What happens with the Cr/Y layer at the grain boundaries when the sample cools down?
This can be investigated with LEIS, too, but some attention has to be paid to the unavoidable sputtering by the He$^{+}$ ions used for the analysis.
Sputtering by the probing ion beam is of no concern for the previous analysis of the thermally driven segregation behaviour, but here it needs to be discriminated against a possible de-segregation.
In order to keep the (preferential) sputtering and thus the surface modification during the measurements as low as possible, the experiment is conducted in the following way:
The sample is cleaned by Kr sputtering and heated to the desired temperature of \SI{1000}{K}.
It is kept at this temperature for sufficient time (about \SI{1}{h}) to allow the segregation to reach an equilibrium.
The time for taking a LEIS spectrum is minimised by (i) reducing the scanned energy range to the Cr, Y, and W peaks (neglecting the small O peak) and (ii) shortening the counting time per energy channel to \SI{4000}{ms}.
Both measures reduce the total time for recording a LEIS spectrum to less than \SI{5}{min}.
With a probing ion beam flux $J = \SI{1e16}{m^{-2}.s^{-1}}$ the maximum Cr sputtering by \SI{1}{keV} He$^{+}$ ions is less than 0.06 monolayers per measurement.
During the waiting time between measurements, when the sample temperature is reduced and stabilises at a lower value, the probing ion beam is blanked by increasing the electric potential of the einzel lens in front of the sample.
Four measurements are recorded and the respective surface concentrations of Cr, Y, and W are shown in figure \ref{fig-cooling}.
The temporal order of the measurement is from right to left, i.e. the measurement at the highest temperature is the first one and the measurement at the lowest temperature is the last one in this series.
In the figure the times at which the spectra are measured (with respect to the first measurement labelled as $t=0$) are indicated.
When going from a higher to a lower temperature the rates of temperature decay are due to radiation losses only and are rather low, no auxiliary sample cooling is applied.
From the time trace of sample temperature versus time one can derive that at the highest temperature the decay rate is less than \SI{0.7}{Ks^{-1}}, and at the lowest temperature the decay rate is smaller than \SI{0.1}{Ks^{-1}}.
The Cr surface concentration stays fairly constant, whereas the Y surface density decays to about half.
The W concentration stays almost zero with the exception of the last measurement which shows a slight increase.
The measurement shows that after cool-down of the sample the segregated Cr layer remains on the surface.
The small increase of the W surface density visible when the sample temperature falls below \SI{400}{K} is likely due to the residual sputtering from the LEIS ion beam which preferentially removes Cr and Y atoms from the surface which are not replenished due to the small segregation at low temperature.
Within the time of the measurement ($>$\SI{2}{h}) no de-segregation is visible.
This behaviour could result in a chromium oxide layer which might explain the enhanced deuterium retention found in W-10Cr-0.5Y samples which were outgassed at even higher temperatures prior to a study of the hydrogen isotope retention \cite{mai19}.

\begin{figure}[ht]
\resizebox{\figurewidth}{!}{\includegraphics{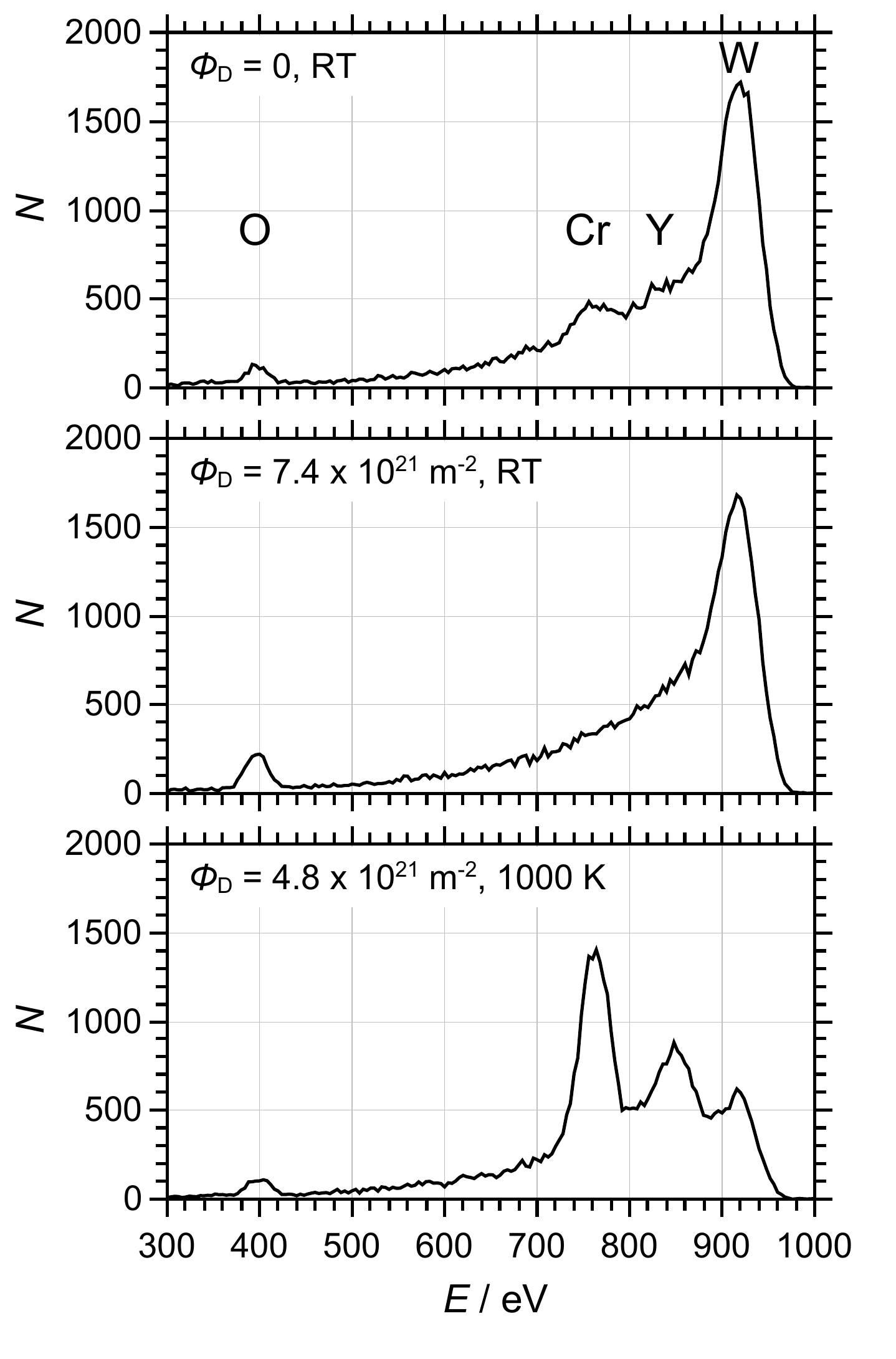}}
\caption{\label{fig-prefsput}\doublespacing{}
LEIS spectra showing the preferential sputtering of the lighter alloy constituents.
The labels indicate the species which cause the respective scattering peaks.
The spectrum after sputter cleaning with Kr is plotted at the top.
The plot in the middle shows the spectrum after D sputtering with a fluence of \SI{7.4e21}{m^{-2}} at RT.
The plot at the bottom displays the spectrum after a D sputter fluence of \SI{4.8e21}{m^{-2}} at a sample temperature of \SI{1000}{K}.}
\end{figure}

The preferential sputtering of the WCrY alloy is investigated with impact of \SI{250}{eV} deuterons.
At this energy there is considerable sputtering of Cr, and almost no erosion of W.
After preparing the sample in the usual way by Kr sputtering an initial LEIS spectrum is measured.
The first preferential sputtering cycle is done at RT with a D fluence of \SI{7.4e21}{m^{-2}}.
The corresponding LEIS spectra are displayed in figure \ref{fig-prefsput}.
The initial spectrum without D sputtering (grey curve) is similar to the spectrum shown in figure \ref{fig-rt-1000k}.
After D bombardment the spectrum exhibits no visible peaks from scattering on Cr or Y.
The slight increase of the O peak at \SI{400}{eV} originates from residual impurities in the D$_{2}$ gas which could not be completely removed by the liquid nitrogen trap\footnote{The gas can contains up to \SI{50}{ppm} O$_{2}$ which, if not reduced by the LN$_{2}$ trap, would contribute by \SI{0.1}{\percent} to the Cr sputtering and about \SI{10}{\percent} to the W sputtering. However, this would reduce the effect of preferential sputtering what is obviously not observed in the measurement.}.
The same preferential sputtering investigation is repeated with the sample at a temperature of \SI{1000}{K} (broken line) using a D fluence of \SI{4.8e21}{m^{-2}}.
The strong segregation of Cr and Y at this sample temperature prevents the complete removal of Cr and Y from the surface and does not result in a full W coverage as it is observed at RT.
Under continuous sputter erosion the lighter segregating alloy constituents Cr and Y are constantly eroded, but at the same time the surface concentrations reach an equilibrium due to the flux of segregating atoms towards the surface.
This can be quantified in the following way:
The flux of sputtering D atoms with impact energy $E$ onto the surface is $J_{\rm D}$.
When e.g. Cr covers a surface fraction $c_{\rm{Cr}}$ and the sputter yield is denoted by $Y_{\rm{Cr}}(E)$, the flux of eroded Cr atoms is
\begin{equation}
 J_{\rm{Cr}}(T) = c_{\rm{Cr}}(T,J_{\rm D}) \times Y_{\rm{Cr}}(E) \times J_{\rm D}.
\label{eq-erosion}
\end{equation}
With $J_{\rm D} = \SI{1e18}{m^{-2}s^{-1}}$, $c_{\rm{Cr}} = 0.55$ (this value of surface coverage corresponds to the \SI{1000}{K} measurement, dashed line in figure \ref{fig-rt-1000k}), and the sputter yield at an energy of \SI{250}{eV} $Y_{\rm{Cr}} = 0.04$, the flux of eroded Cr atoms is \SI{2e16}{m^{-2}s^{-1}}.
Note that the sputtered Cr flux in equation \ref{eq-erosion} in the equilibrium state (${\rm d}c/{\rm d}t = 0$) is equal to the flux of Cr atoms which flows from the bulk to the surface due to thermal segregation.
This flux depends only on the temperature and on the equilibrium surface coverage and is not specific to the used ion beam setup.
However, although equation \ref{eq-erosion} looks rather simple, the surface coverage $c_{\rm{Cr}}(T,J_{\rm D})$ which develops for specific irradiation conditions is non-linear.

It seems still possible to make a rough estimate for the behaviour of a WCrY wall cladding in a future fusion reactor.
Using a typical D flux of \SI{1e20}{m^{-2}s^{-1}} onto the first wall (figure 4 in \cite{beh03}) and an assumed impact energy of \SI{50}{eV}, the sputter yield for (elemental) Cr is \num{1e-4} \cite{eck07}.
The product $Y \times J$ equals \SI{1e16}{m^{-2}s^{-1}} which is even lower than (but not very different to) the value obtained in the present ion beam setup.
Assuming a resultant Cr surface fraction of e.g. \num{0.5}, which can always be complied at a certain temperature (above \SI{700}{K} where sufficient segregation occurs), the sputtered Cr flux will be \SI{5e15}{m^{-2}s^{-1}}, i.e. the flux is of the same order as in the present experiment.
The WCrY first wall cladding in a fusion reactor will be rather thin in order not to weaken the tritium breeding ratio too much.
Taking a thickness of \SI{2}{mm}, the density of Cr atoms will be \SI{4e25}{m^{-2}}.
This number has to be compared to the loss rate of segregated Cr atoms due to sputter erosion which is much smaller and would lead to a reduction of only \SI{0.4}{\percent} of the Cr atoms per operational year.
Re-deposition of sputtered Cr due to the large magnetic field in a fusion device might lead to a further reduction of the loss rate.
Even eroding D fluxes larger by one order of magnitude might not be critical at all.

\newpage{}
\section{Summary and Conclusion}
The elemental surface composition of self-passivating WCrY samples has been analysed with LEIS at various temperatures up to \SI{1000}{K}.
Strong Cr and Y segregation on the surface occurs above \SI{700}{K} and W atoms in the topmost surface layer are not detectable at the highest temperature but are still present in the layers beneath the topmost surface layer.
Below \SI{700}{K} a much weaker surface density increase of Cr and Y is observed which might not necessarily be attributed to segregation but could result from the recovery of the equilibrium surface concentrations after the sputter cleaning of the sample.
According to the W-Cr phase diagram (NB: a W-Cr-Y phase diagram is not yet available) the alloy with \SI{31}{\percent} Cr at a temperature of \SI{1000}{K} is located in the miscibility gap which consists of a mixture of Cr-rich and W-rich solid solutions.
The observed segregation can be attributed to grain boundary segregation of these phases.
The segregation measurements at the various temperatures are analysed with a Langmuir-McLean relation and yield segregation enthalpies of \SI{0.7}{eV} for Cr and \SI{0.3}{eV} for Y in the temperature range above \SI{700}{K}.

D$_{2}^{+}$ sputtering with \SI{250}{eV} per deuteron results in preferential sputtering and almost complete Cr depletion from the surface at RT.
At an increased sample temperature of \SI{800}{K} or \SI{1000}{K} the D sputtering does not completely remove the Cr atoms from the surface.
Instead, the strong thermal segregation still yields an increase of the Cr surface concentration.
The equilibrium between sputter erosion and segregation allows to determine the flux of Cr atoms towards the surface which are eroded and lost from the bulk.

Although the Cr segregation prevents the build up of a complete W layer, no serious issue which might question the application of the WCrY alloy for the intended purpose has emerged by this finding.
The enhancement of Cr erosion is rather weak and and extrapolation for the conditions in a fusion reactor yields Cr losses less than \SI{1}{\percent} per operational year.
In case of a loss-of-coolant accident the Cr enrichment on the surface might even turn out to be beneficial because it accelerates the formation of the protecting chromium oxide layer.

\newpage{}
\section*{Acknowledgements}
We thank Mr. Albert Hiller and Mr. Roland B\"ar for technical assistance and maintenance of the apparatus, and Mrs. Beatrix G\"oths for polishing the WCrY sample.

Helpful discussion on the role of the chromium oxide layer for hydrogen retention with Dr. Hans Maier (IPP Garching) is very much acknowledged.

This work has been carried out within the framework of the EUROfusion Consortium and has received funding from the Euratom research and training programme 2014-2018 and 2019-2020 under grant agreement No 633053. The views and opinions expressed herein do not necessarily reflect those of the European Commission.


\newpage{}
\renewcommand{\arraystretch}{2.0}
\setlength{\tabcolsep}{20pt}
\begin{table}[h]
\begin{tabular}[h]{ l r r r }
\hline
       & W    & Cr   & Y  \\ \hline
weight\% & 88.0 & 11.4 & 0.6 \\
atom\% & 67.9 & 31.1 & 1.0 \\
\hline
\end{tabular}
\caption{\label{tab-wcry}\doublespacing{}
Bulk composition of the WCrY alloy sample.}
\end{table}


\newpage

\newpage

\newpage

\newpage

\newpage

\newpage

\newpage

\end{document}